\documentclass[12pt,preprint]{aastex}

\usepackage{amsmath}
\usepackage{color}

\shorttitle{A search for sub-second variability toward 3C~84}
\shortauthors{Hales et al.}

\begin{document}

\title{
A search for sub-second radio variability predicted to arise
\\
toward 3C~84 from intergalactic dispersion
}

\author{C.~A. Hales}
\affil{National Radio Astronomy Observatory, P.O. Box 0, Socorro, NM 87801, USA}

\author{W. Max-Moerbeck}
\affil{National Radio Astronomy Observatory, P.O. Box 0, Socorro, NM 87801, USA}
\affil{Max-Planck-Institut f\"{u}r Radioastronomie, Auf dem H\"{u}gel 69, D-53121 Bonn, Germany}

\author{D.~A. Roshi}
\affil{National Radio Astronomy Observatory, Charlottesville \& Green Bank,
       520 Edgemont Road, Charlottesville, VA 22903, USA}

\and

\author{M. P. Rupen}
\affil{National Research Council of Canada, Herzberg Astronomy and Astrophysics,
       Dominion Radio Astrophysical Observatory, P.O. Box 248, Penticton, BC V2A 6J9, Canada
       \vspace{-1mm}}

\begin{abstract}
\vspace{-8mm}
We empirically evaluate the scheme proposed by \citet{2013ApJ...763L..44L} in which the
light curve of a time-steady radio source is predicted to exhibit increased variability
on a characteristic timescale set by the sightline's electron column density. Application
to extragalactic sources is of significant appeal as it would enable a unique and reliable
probe of cosmic baryons. We examine temporal power spectra for 3C~84 observed at
1.7~GHz with the Karl G. Jansky Very Large Array and the Robert C. Byrd Green Bank Telescope.
These data constrain the ratio between standard deviation and mean intensity for 3C~84 to less
than 0.05\% at temporal frequencies ranging between 0.1--200~Hz. This limit is 3 orders of
magnitude below the variability predicted by \citet{2013ApJ...763L..44L} and is in accord
with theoretical arguments presented by \citet{2014MNRAS.440.3613H} rebutting electron
density dependence. We identify other spectral features in the data consistent with the
slow solar wind, a coronal mass ejection, and the ionosphere.
\end{abstract}


\section{Introduction}

Approximately 30\% of all baryons in the low redshift ($z$~{\footnotesize $\lesssim$}~1)
Universe remain unaccounted for observationally. Simulations predict that the bulk, if not
all, of these `missing baryons' reside at shock-heated temperatures between $10^5$~K to $10^7$~K
in the diffuse warm-hot intergalactic medium (WHIM) of the mildly non-linear cosmic web
\citep{1999ApJ...514....1C,2001ApJ...552..473D,2007ARA&A..45..221B,2012ApJ...759...23S}.
Similar temperatures may also be present in voids
\citep{2012ApJ...752...23C,2014ApJ...797..110C,2015MNRAS.448.3405M}.
Detecting gas under these conditions is challenging. Emission and absorption based probes
such as the Sunyaev-Zel'dovich effect and spectral line diagnostics, in combination with
gravitational lensing and cross-correlation studies, are becoming increasingly
sensitive to the WHIM \citep[e.g.][]{2009ApJ...695.1351B,2010MNRAS.402....2B,
2012Natur.487..202D,2014MNRAS.445..460G,2015ApJ...806..113G,2016MNRAS.455.2662T}.
However, these tracers are complicated by degeneracies with physical conditions
including temperature, density structure, metallicity, magnetic field, and ionization
state.

An alternate tracer that provides equal sensitivity to every free electron along the
line of sight, irrespective of gas condition, is the dispersion measure,
$\textrm{DM} = \int n_e(z) \, \textrm{d}l$, where $n_e(z)$ is the electron density as a
function of redshift and d$l$ is differential proper distance. Radio band observations
of the frequency-dependent arrival times from an impulsive event, such as a pulsar pulse,
can be used to infer the DM. Given that the Universe is essentially fully ionized at
$z<2.5$ \citep{2009ApJ...694..842M}, intergalactic DMs offer excellent prospects for
mapping baryons throughout large scale structure including the WHIM. The challenge
to date, however, has been the lack of suitable extragalactic targets. Pulsar DMs have
been used to successfully map ionized gas throughout the Milky Way \citep{2002astro.ph..7156C},
but pulsars are too faint to be detected beyond $\sim100$~kpc. Variable radio emission
from quasars and gamma-ray bursts has been proposed
\citep[e.g.][]{1965PhRvL..14.1007H,1993ApJ...417L..25P}, but the lightcurves for these sources
do not contain suitably sharp features against which to measure dispersive delays.

Excitingly, this situation may have improved with the discovery of a new population.
Fast radio bursts (FRBs) are bright, isolated, highly dispersed, millisecond-duration flashes
of unknown astrophysical origin, discovered with single-dish radio telescopes and inferred to
have a high all-sky event rate \citep{2007Sci...318..777L,2013Sci...341...53T,2015MNRAS.447.2852K}.
Their large dispersion measures are consistent with a population of standard candles
observable to $z$~{\footnotesize $\lesssim$}~1 \citep{2014ApJ...785L..26L,2015MNRAS.451.4277D},
though less distant candidates located within $z$~{\footnotesize $\lesssim$}~0.1
\citep{2016MNRAS.457..232C,2015ApJ...807..179P} or even purely within our Galaxy \citep{2015MNRAS.454.2183M}
may also explain their observed properties. \citet{2016Natur.530..453K}
recently claimed the first localization of an FRB, using the detection of a fading radio transient to identify
the host as an elliptical galaxy at at $z=0.492$. However, their FRB--afterglow association has been
strongly disputed as likely arising from scintillation of an unrelated background active galactic nucleus
\citep{2016arXiv160208434W,2016arXiv160304880A}. \citet{2016Natur.531..202S} subsequently report
in a separate work the discovery of a repeating FRB, supporting an origin in a nearby extragalactic
($z$~{\footnotesize $\lesssim$}~0.1) neutron star. \citet{2016Natur.531..202S} note that the non-detection
of repeating bursts from other FRBs may indicate the existence of more than one mechanism for producing FRBs.
If an additional class of FRB hosts can be ranged to cosmological distances
\citep[e.g.][]{2015MNRAS.451L..75F,2015arXiv151103615M}, their DMs can be used to address a range of science.
Examples include mapping the three-dimensional distribution and clustering of cosmic baryons
\citep{2015PhRvL.115l1301M}, quantifying the accretion onto and feedback within galactic halos
\citep{2014ApJ...780L..33M}, measuring the intergalactic baryon mass fraction \citep{2014ApJ...783L..35D},
constraining the dark energy equation of state \citep{2014ApJ...788..189G,2014PhRvD..89j7303Z},
providing insights into turbulence in the intergalactic medium and the origin and evolution of
cosmic magnetism by helping to unravel the degeneracy between electron density and magnetic field
strength in Faraday rotation measures \citep{2009MNRAS.392.1008D,2012SSRv..166....1R,2016arXiv160203235A},
and possibly probing He~II reionization at $z\sim3$ if a population of FRBs is present up to
these redshifts \citep{2014ApJ...797...71Z}.

Access to the science above could also be gained through the development of new observational
techniques. Pursuing this strategy, \citet{2013ApJ...763L..44L} and \citet{2013MNRAS.433.2275L}
raised the revolutionary prospect of measuring DMs toward time-steady rather than impulsive
sources. These works predict the appearance of fluctuations in the light from
distant radio sources on timescales of milliseconds, with the characteristic timescale for each
source proportional to its DM. If correct, these schemes would enable the cosmic baryon science
above to be addressed using known populations, such as radio quasars with confirmed ranges out
to and beyond $z\sim4$. However, a comprehensive theoretical rebuttal was subsequently presented by
\citet{2014MNRAS.440.3613H}, ruling out any expected dependence between DM and the statistical
properties of radiation received from a time-steady radio source. \citet{2014MNRAS.440.3613H}
derive the 2- and 4-point correlation of the received electric field (including examination of the
quantum nature of the electromagnetic field for the latter), corresponding to the \citet{2013MNRAS.433.2275L}
and \citet{2013ApJ...763L..44L} techniques respectively. They demonstrate under general principles
that the former is insensitive to DM, and the latter has extremely weak dependence due to the
central limit theorem for a large number of incoherently emitting electrons.

In this work we present an independent empirical evaluation of the technique proposed
by \citet{2013ApJ...763L..44L}; we do not evaluate the technique proposed by \citet{2013MNRAS.433.2275L}
due to challenging sensitivity criteria. While the theoretical results of 
\citet{2014MNRAS.440.3613H} convincingly rule out the need for empirical examination, we present
the findings of our customized observations here as an independent approach and as a demonstration of the
ancillary science available through fast cadence radio observations. Section~\ref{sec2} describes our
experiment design. Section~\ref{sec3} describes our radio observations and data reduction. Results are
presented in Section~\ref{sec4} with discussion in Section~\ref{sec5}. We conclude in Section~\ref{sec6}.

\section{Experiment Design}\label{sec2}

The techniques proposed by \citet{2013ApJ...763L..44L} and \citet{2013MNRAS.433.2275L} are based on the
plasma dispersion effect whereby photon wave packets undergo broadening in their envelope width and a
drift in their carrier frequency as they traverse vast spans of the ionized intergalactic medium.
Both works argue that the statistical properties of radiation received from a time-steady incoherent
source will be altered in the presence of plasma dispersion, provided the radiation is composed of many
temporally overlapping wave packets (e.g. synchrotron). \citet{2013ApJ...763L..44L} predict that, in
the case of no intervening plasma, strong intensity fluctuations will be observed with standard deviation
half\footnote{In a follow-up paper, \citet{2013ApJ...778...73L} give the standard deviation as the mean
intensity of the source.} the mean intensity of the source, $\sigma_I=\bar{I}/2$, and correlation timescale
given by the inverse of the observation bandwidth, $\tau_c=(2\pi\Delta\nu)^{-1}$. When dispersive plasma
is present, they predict intensity fluctuations spread over a range of timescales. Fourier power is
predicted to rise monotonically from $\tau_c$ until exhibiting the full plasma-free variability level
($\sigma_I=\bar{I}/2$) at a `stretched envelope' timescale given by
\begin{equation}
	\tau_\textrm{\scriptsize{env}} = \sqrt{2} \, \tau_c \left\{
	1+\left[
	5\times10^6 \left(\frac{\Delta\nu/\nu_0}{10^{-2}}\right)^2
	\bigg(\frac{\nu_0}{\textrm{1 GHz}}\bigg)^{-1}
	\left(\frac{\textrm{DM}}{\textrm{$10^3$ pc cm}^{-3}}\right)
	\right]^2 \right\}^{1/2} \; \textrm{[s]}
	\label{eqn:lieu}
\end{equation}
for central observing frequency $\nu_0$. Similarly, \citet{2013MNRAS.433.2275L} predict that plasma
dispersion will cause a peak to appear in the two-point correlation of the electric field between two
nearby frequencies.

To empirically test the behavior predicted by \citet{2013ApJ...763L..44L}, we seek to analyze
radio-band light curves of a target with known DM. To minimize any possible confusion with rapid
variability caused by ionospheric, interplanetary, or interstellar scintillation
\citep[e.g.][]{1950Natur.165..422S,clarke,1964Natur.203.1214H,2002Natur.415...57D},
we require sampling on sub-second timescales \citep[intergalactic scintillation timescales
are many orders of magnitude larger and thus negligible here;][]{2013MNRAS.434.3293P}. This in turn
requires a strong radio source to ensure high signal to noise within short integration periods.

For this purpose we select 3C~84, the radio source associated with the giant elliptical
galaxy NGC~1275, located at $z=0.018$ at the center of the brightest X-ray cluster in
the sky, Perseus. 3C~84 is one of the strongest and most compact radio sources in the sky
\citep{1976Natur.259...17P}. While 3C~84 is known to be variable on timescales of years
\citep{1966ApJ...144..843D,2009ApJ...699...31A}, we do not expect any intrinsic
sub-second variability because the light crossing time for its $3.4\times10^8$~M$_\sun$
black hole \citep{2005MNRAS.359..755W} is $\sim1$~hour (though cf. plausible rapid variability
due to opacity variations described by \citealt{1979ApJ...228...27M} and subhorizon-scale
black hole lightning described by \citealt{2014Sci...346.1080A}).

We calculate the total DM along the line of sight to 3C~84 as follows. The contribution
from the intracluster medium within Perseus is estimated to be 2700~pc~cm$^{-3}$.
This is obtained by integrating the electron density profile inferred from X-ray observations.
We model this profile with a mean value 0.3~cm$^{-3}$ within radius $r<2$~kpc \citep{2006MNRAS.368.1500T}
that declines beyond as $0.31(r/\textrm{kpc})^{-0.86}$ out to $r=200$~kpc
\citep{1981ApJ...248...47F,2006MNRAS.366..417F}. The foreground Milky Way contribution is
approximately 100~pc~cm$^{-3}$ \citep{2002astro.ph..7156C}. The total DM is therefore
approximately 2800~pc~cm$^{-3}$. We place overly conservative lower and upper limits on this total
DM of 1500~pc~cm$^{-3}$ and 6000~pc~cm$^{-3}$, respectively. These limits are intended to account
for all plausible uncertainties, including additional DM contributions due to the
intergalactic path between Perseus and our Galaxy
\citep[likely $<100$~pc~cm$^{-3}$;][]{2003ApJ...598L..79I,2014ApJ...780L..33M},
contributions local to NGC~1275, and differences between our line of sight and the
in-sky X-ray profile used to estimate the dominant intracluster medium contribution
\citep[the halo geometry is likely non-spherical;][]{1990MNRAS.246..477P}.

To mitigate potential instrumental systematics, we pursue observations with
two independent facilities: the Karl G. Jansky Very Large Array (VLA) and the Robert C.
Byrd Green Bank Telescope (GBT). The maximum throughput of the VLA's WIDAR correlator
is approximately 1~TB~hr$^{-1}$. Access to this data rate is only available through
non-standard correlator configuations. The special mode commissioned by \citet{2015ApJ...807...16L}
for FRB studies provides 5~ms time resolution for $2\times128$~MHz spectral windows at L-band. We
make use of this mode here as it is well matched to our experiment. We pursue observations with the
GBT's VEGAS backend in Mode 1 with 2.5~ms sampling.

We tune our observing setup as follows. We set the requirement $\tau_\textrm{\scriptsize{env}}<40$~Hz
so that Fourier behavior on faster timescales (toward $\tau_c$) can be gleaned up to the 100~Hz Nyquist
frequency accessible with the VLA (up to 200~Hz is accessible with 2.5~ms sampling at the GBT).
We seek observing frequencies near 1760~MHz to avoid typical radio frequency interference (RFI) at
the VLA and GBT sites. To satisfy these requirements with the DM range predicted above, according to
Equation~(\ref{eqn:lieu}) we require $\Delta\nu=8$~MHz ($\tau_c=20$~ns). For this setup, Equation~(\ref{eqn:lieu})
therefore predicts $\tau_\textrm{\scriptsize{env}}$ in the range $10-40$~Hz, with 22~Hz corresponding
to $\textrm{DM}=2800$~pc~cm$^{-3}$. The largest bandwidth for which $\tau_\textrm{\scriptsize{env}}$
will remain on sub-second timescales is $\Delta\nu=90$~MHz ($\tau_c=2$~ns). For this bandwidth,
Equation~(\ref{eqn:lieu}) predicts $\tau_\textrm{\scriptsize{env}}=1-4$~Hz, with 2~Hz corresponding
to $\textrm{DM}=2800$~pc~cm$^{-3}$.

We do not investigate the predictions from \citet{2013MNRAS.433.2275L} in this work. This is due to
challenges in meeting signal to noise requirements within narrow channel bandwidths at high time resolution,
which can only be overcome by combining channels in a customized filter bank (similar to those used
in pulsar astronomy). While possible, we did not ultimately pursue this strategy due to the findings from
\citet{2014MNRAS.440.3613H} and the conclusions reported in this work for the \citet{2013ApJ...763L..44L}
experiment.

\section{Observations and Data Reduction}\label{sec3}

Observations of 3C~84 are summarized in Table~\ref{tbl-obs}. The on-target duration was approximately 5~min
per epoch. We observed the flux density calibrator 3C~48 in each session.
\begin{table}
\begin{center}
\caption{Observations of 3C~84.\vspace{2mm}\label{tbl-obs}}
\begin{tabular}{ccccc}
\tableline\tableline
Facility & Configuration & Project Code & Epoch (UTC) & Solar Elongation ($^\circ$)\\
\tableline
GBT & -- & 14A-418 & 2014 Apr 11 14:18 & 43\\
VLA & A & 14A-409 & 2014 May 15 15:28 & 23\\
VLA & C & 14B-503 & 2015 Jan 02 00:10 & 133\\
\tableline
\end{tabular}
\end{center}
\end{table}

We initially performed simultaneous observations with the VLA and GBT on 2014 April 11.
Unfortunately, these VLA data were corrupted due to a correlator error. Successful observations
with the VLA were subsequently obtained on 2014 May 15. Additional VLA observations were obtained
on 2015 January 2 at large solar elongation to discriminate the effects of interplanetary
scintillation in the low-elongation 2014 VLA and GBT data.

Specifics of the observations and data reduction procedures for each facility are detailed as follows.

\subsection{Very Large Array}

VLA observations were performed using the correlator mode developed by \citet{2015ApJ...807...16L},
delivering dual circular polarization products in two spectral windows at 5~ms time resolution.
We centered these spectral windows at 1410 and 1760~MHz, each spanning $64\times2$~MHz channels.
Results from the 1410~MHz window are consistent with the 1760~MHz window and will not be described
further in this work (the dual-window setup was originally intended to evaluate the proposed
frequency dependence for $\tau_\textrm{\scriptsize{env}}$).

The data were reduced using standard tasks in version 4.3.0 of the CASA package \citep{2007ASPC..376..127M}.
Hanning smoothing was applied. Only a few percent of the data were found to be affected by RFI. These were
manually identified and flagged. Our final results are unaffected if the Hanning smoothing step is removed,
due to the minimal presence of RFI in the data. Flux density calibration was bootstrapped
against 3C~48, adopting the most recent 2012 value from the \citet{2013ApJS..204...19P} standard.
Phase calibration was performed for each 5~ms integration for both 3C~48 and 3C~84 to verify
data fidelity and to remove positional shifts due to scintillation in preparation for image-based
analysis. The solar elongations for 3C~48 at the 2014 and 2015 epochs were $28^\circ$ and $112^\circ$,
respectively.

We extracted 120~s of continuous 3C~84 sampling within two bandwidths, spanning 8~MHz and 90~MHz,
from both the 2014 and 2015 data. Central frequencies for the 8~MHz bands were selected to ensure they
did not exhibit any RFI over the duration of the observations. The 90~MHz bands were selected to
minimize the inclusion of any spectral channels where RFI had been flagged. These selections resulted in
slightly different central frequencies for the two bandwidths. The central frequencies for the 8~MHz
bandwidth data are 1765~MHz and 1761~MHz for the respective consecutive epochs. The corresponding central
frequencies for the 90~MHz bandwidth data are 1759~MHz and 1757~MHz. Differences in predicted
$\tau_\textrm{\scriptsize{env}}$ are negligible. Light curves were constructed by imaging each
5~ms integration and measuring 3C~84's flux density using the peak surface brightness. The observed
noise per integration is approximately 77~mJy~beam$^{-1}$ for the 8~MHz data and 23~mJy~beam$^{-1}$
for the 90~MHz data, consistent with theoretical predictions.

Extensive diagnostic checks were performed throughout this process to search for and identify any
spurious instrumentally-induced signatures, for example a tone identified at 4~Hz that was traced
to a memory failure in the delay module on the station board for antenna EA09.

\subsection{Green Bank Telescope}

GBT observations were performed with the VEGAS\footnote{http://www.gb.nrao.edu/vegas/modes} backend
using Mode 1, delivering dual linear polarization data spanning the frequency range 0.8--2.3~GHz split into
1024 spectral channels. The maximum useable bandwidth at L-band is limited by the feed to 650~MHz. Of this,
we restricted our calibration to the range 1704--1814~MHz and focused our experimental analysis on two
bands spanning 6 ($\approx8$~MHz) and 62 ($\approx90$~MHz) channels centered at approximately 1736~MHz and 1759~MHz,
respectively. We selected the central frequency of the 8~MHz band to be as close as possible to 1760~MHz
while avoiding any RFI. Differences in predicted $\tau_\textrm{\scriptsize{env}}$ due to the different
frequency and bandwidth selections, and compared to the VLA observations, are negligible.

We performed two types of observations. First, we observed 3C~48 and 3C~84 at 1~s time resolution with noise
diode switching enabled (1~s period with power $\sim10\%$ of the system temperature). We then repeated these
observations using 2.5~ms time resolution, but with noise diode switching turned off to prevent data blanking.
In both cases we employed position switching with reference points located 45\arcmin\ north of 3C~48 and 45\arcmin\
south of 3C~84. These positions are devoid of sources in the 1.4~GHz NRAO VLA Sky Survey \citep{1998AJ....115.1693C}
and exhibit negligible-gradient background 408~MHz brightnesses within approximately 2\% of the target
position values \citep{1982A&AS...47....1H}.

The data were calibrated independently per polarization. The absence of the noise diode signal required a non-standard
calibration which we implemented in Python based on the GBTIDL\footnote{http://gbtidl.nrao.edu} procedures
outlined by \citet{braatz}. Only a few percent of the data in our frequency range of interest were found to be
affected by RFI. These data were manually identified and flagged. We used Equation~(1) from \citet{braatz} to
calculate the time-averaged system temperature, $T_{\rm sys}$, for each of 3C~48 and 3C~84 using the 1~s data. We used
Equation~(2) from \citet{braatz} to calculate the time-averaged antenna temperature as a function of spectral channel,
$T_{\rm ant}(\nu)$, for 3C~48 using the 1~s data. To calculate the antenna temperature as a function of time and
channel for 3C~84 using the 2.5~ms data, $T_{\rm ant,sig}(t,\nu)$, we first calculated the following modified
system temperature to account for the absence of the noise diode signal,
\begin{equation}
    T_{\rm sys}^{\prime} = T_{\rm sys} ~
    \left\langle \frac{(\overline{{\rm sig}} - \overline{{\rm ref}})}{\overline{{\rm ref}}}
    \bigg/ \frac{(\overline{{\rm sig}_{\rm caloff}} - \overline{{\rm ref}_{\rm caloff}})}{\overline{{\rm ref}_{\rm caloff}}}
    \right\rangle \,, \label{eqn:tsys}
\end{equation}
where $\rm{sig} \equiv \rm{sig}(t,\nu)$ and $\rm{ref} \equiv \rm{ref}(t,\nu)$ are position-switched
target and reference data, respectively, bars and angular brackets indicate time and channel averages,
respectively, and the caloff subscripts indicate 1~s data where the calibration noise diode is off.
We then calculated
\begin{equation}
    T_{\rm ant,sig}(t,\nu) = T_{\rm sys}'~\frac{{\rm sig} -
    \overline{{\rm ref}}}{\overline{{\rm ref}}} \,,
    \label{eqn:tant}
\end{equation}
and similarly $T_{\rm ant,ref}(t,\nu)$ for the reference position with sig replaced by ref.
We confirmed this strategy by calculating Equation~(\ref{eqn:tant}) for 3C~48 using the 2.5~ms data. The result was
consistent with the original $T_{\rm ant}$ derived from the 1~s data.

The data for both linear polarizations were then averaged. Corrected antenna temperatures were calibrated to
the most recent 2012 flux density for 3C~48 from the \citet{2013ApJS..204...19P} standard. Channels were
averaged to form our two bands of interest. From these, we extracted 85~s of continuous sampling for 3C~84.
We calculate the noise in our data by integrating the power spectrum for each light curve between
100~Hz and the 200~Hz Nyquist frequency, excluding lower temporal frequencies that are dominated by real signal
(see Fig.~\ref{fig2} presented shortly). The observed noise per sample is approximately 220~mJy~beam$^{-1}$
for the 8~MHz data and 66~mJy~beam$^{-1}$ for the 90~MHz data, consistent with theoretical predictions.

We examined the 1~s reference position data for evidence of gain fluctuations in the receiver that would
appear as flicker noise \citep{mason}. If present, such fluctuations could potentially be misidentified as
having a non-instrumental origin. We found no evidence for flicker noise above a temporal frequency of 0.07~Hz.
We found tentative evidence for increased spectral power below 0.07~Hz. This temporal frequency is much
lower than the timescales of interest in this work. Additionally, our primary interest in this work regards
the 2.5~ms data for which the thermal noise is greater. To avoid any possible confusion, in this work we
only focus on temporal frequencies greater than 0.1~Hz.

\section{Results}\label{sec4}

\begin{figure}
\includegraphics[angle=-90,width=140mm]{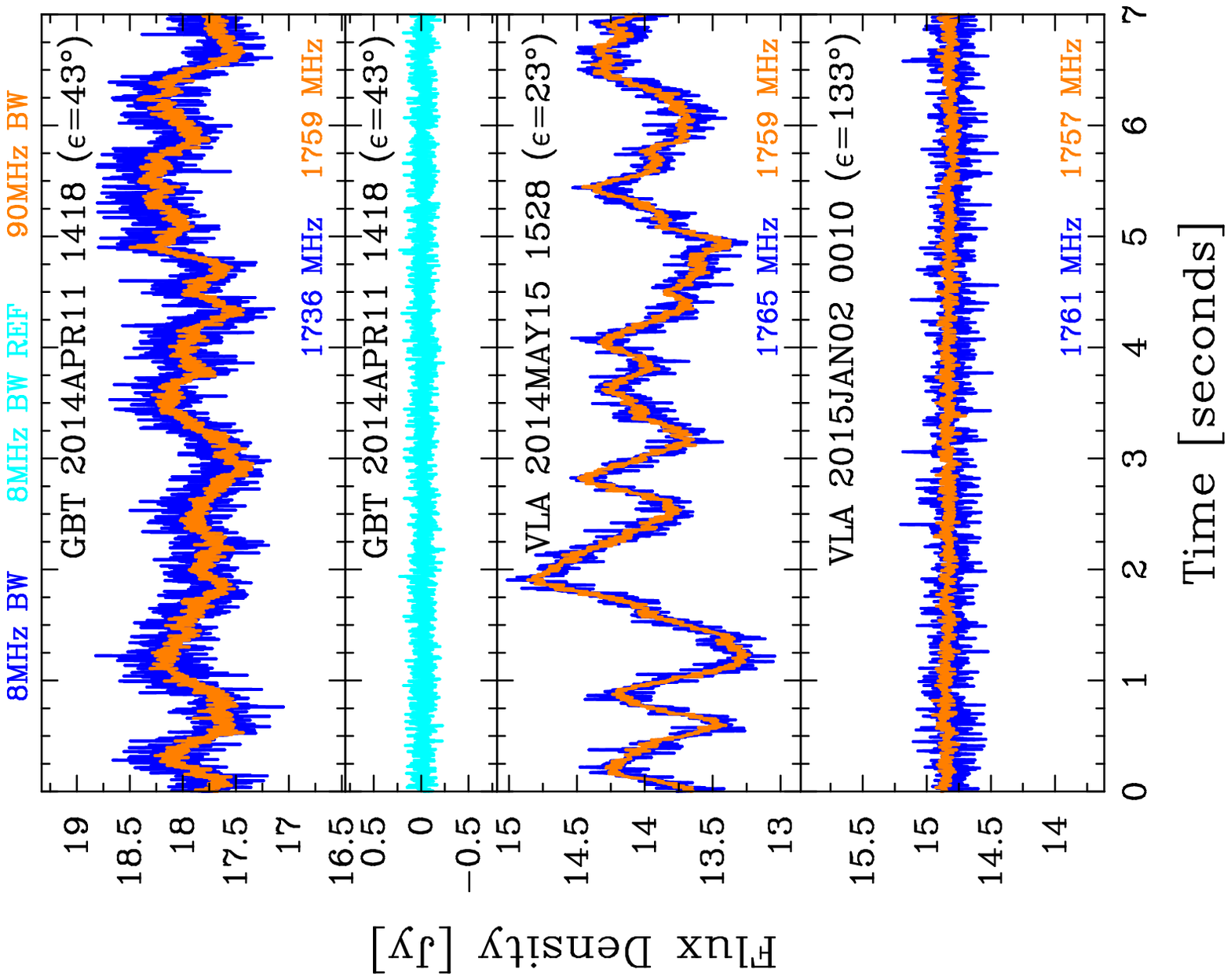}
\caption{
VLA and GBT light curves of 3C~84, spanning representative 7~s portions from observations at
3 epochs. The central frequencies of the 8~MHz (dark blue) and 90~MHz (orange) bandwidth data
are indicated in each panel. The 8~MHz data for the GBT 3C~84 reference position is shown in
the second panel (cyan). The range in each panel, except the second, spans $\pm8\%$ about
the mean flux density of the 8~MHz data.
\label{fig1}}
\end{figure}

The VLA and GBT light curves of 3C~84 are displayed in Fig.~\ref{fig1}.
The mean flux densities of the 8~MHz and 90~MHz bandwidth data differ at each given epoch due to their
slightly different central frequencies and the spectrum of 3C~84. Our data indicate that this spectrum is
falling\footnote{Historically this has not always been the case at our observing frequency
\citep[e.g.][]{1980SvAL....6..322B}.} as a function of frequency \citep[e.g.][]{2009ASPC..402..106N}.
Comparing between epochs, the mean flux densities of the GBT data are higher than the VLA data, and there
is a slight increase between the 2014 and 2015 VLA data. The former difference is due to extended structure
about 3C~84 to which our interferometric observations are insensitive; 3C~84 is located within a $5^\prime$
halo at 1.4~GHz \citep{1962ApJS....7..141M,1968MNRAS.138....1R}. The difference between the 2014 and 2015
epochs is caused by two factors. First, the 2015 data were obtained in C configuration, delivering increased
sensitivity to the extended halo compared to the 2014 A configuration data. Second, 15~GHz pointing
calibration data from the Very Long Baseline Array (VLBA) indicate that 3C~84 underwent a flare that
peaked approximately 2014 July, with flux density at our 2015 epoch a few percent higher compared to
the 2014 epoch. A third (opposite) contribution could arise from our adoption of a fixed 2012 flux
density for 3C~48. If the flux density of 3C~48 has continued to decline following the historical trend
shown by \citet{2013ApJS..204...19P}, our flux densities for 3C~84 will be underestimated, though by
no more than about 2\% at the 2015 epoch. The conclusions presented in this work are not affected by
the differences in mean flux densities seen in Fig.~\ref{fig1}.

\begin{figure}
\includegraphics[angle=-90,width=140mm]{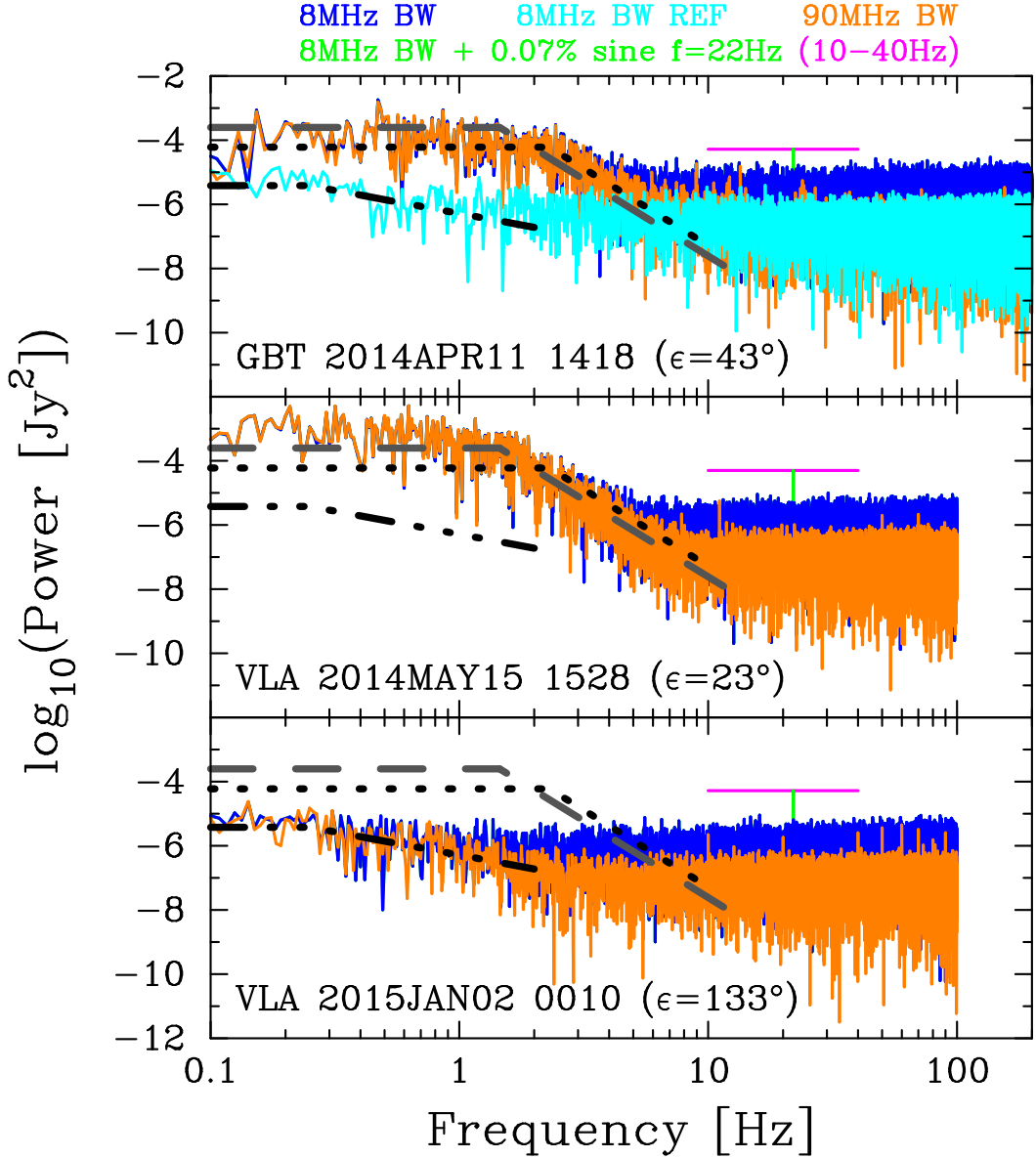}
\caption{
Temporal power spectra of light curves for 3C~84 at 3 epochs. The layout and color scheme
matches Fig.~\ref{fig1}. The 5~ms and 2.5~ms sampling of the VLA and GBT data permit examination
to 100~Hz and 200~Hz, respectively. The dotted, dashed, and dot-dashed curves are informal fits
characterizing the rise and plateau of spectra with decreasing frequency and are identical in
each panel. See \S~\ref{sec4} and \S~\ref{sec5:subHz} for details. The green curve is calculated
from the 8~MHz light curve at each epoch, but with a sinusoid added with period 22~Hz, mean
intensity given by the mean of the original 8~MHz data, and standard deviation of fluctuations
0.05\% (or wave amplitude 0.07\%) of the mean intensity. The purple line indicates the power
that would result for a similarly injected sinusoid with period bewteen 10--40~Hz.
\label{fig2}}
\end{figure}

Fig.~\ref{fig2} displays power spectra\footnote{We calculate power spectra
as the modulus squared of the discrete Fourier transform, normalized such that the sum over positive frequencies
is equal to the variance of the input light curve.} for the light curves sampled in Fig.~\ref{fig1}.
From high to low temporal frequencies, the spectra are characterized by flat power levels that
rise to a plateau. The observed power at high frequencies is limited by sensitivity, as evidenced
by expected lower power levels for the 90~MHz bandwidth data compared to the 8~MHz data and similarly
between the VLA and GBT data. At lower frequencies as the spectra rise, correspondence between
the 8~MHz and 90~MHz data improves, indicating Fourier power dominated by signal and not noise.
The difference between the GBT target and reference data at low frequencies indicates that the
increased Fourier power is real (for clarity, only the 8~MHz reference data is shown
in Fig.'s~\ref{fig1} and \ref{fig2}; the 90~MHz reference data exhibit the same trends).
The difference in power levels between the GBT target and reference 8~MHz data at high
frequencies is consistent with increased noise due to the presence of the $5^\prime$ halo
about 3C~84; the system temperature calculated at the reference position (used for calibration)
is a factor 3 smaller than a similarly calculated system temperature at the position of 3C~84.

The curves displayed in Fig.~\ref{fig2} are informal fits to guide the eye. The dotted and
dashed curves characterize the GBT and 2014 VLA data, respectively, using a power law with
common slope -4.8. The dotted curve flattens at 2.3~Hz while the dashed curve flattens
at 1.5~Hz. These frequencies correspond to the strongest fluctuations seen on $\sim0.5$~s
timescales in Fig.~\ref{fig1}. The dot-dashed curve characterizes both the GBT reference
position data and 2015 VLA data using a power law with slope -1.4 that flattens at 0.25~Hz.

We place a conservative upper limit on the presence of any signals in the 10--40~Hz range by
injecting sine waves with specified amplitude into the 8~MHz light curves, as described in
the caption to Fig.~\ref{fig2}. At all epochs we limit the standard deviation of fluctuations
to less than 0.05\% of the mean intensity. Examining the GBT and 2015 VLA data, the standard
deviation to mean ratio is less than 0.05\% over the full 0.1--200~Hz range. Similar
calculations in the 1--4~Hz range (or indeed over the full 0.1--200~Hz range) for the 90~MHz
light curves yield a consistent result. For reference, the variability
predicted by the theory of \citet{2013ApJ...763L..44L} would appear in Fig.~\ref{fig2} at
power levels near 100~Jy$^2$.

\section{Discussion}\label{sec5}

\subsection{Implications for Proposed Theory}

Our empirical limits on variability in the 1--4~Hz and 10--40~Hz ranges for 3C~84 are 3 orders
of magnitude less than the levels predicted by the theory of \citet{2013ApJ...763L..44L}. We do not
see any evidence for a decline in spectral power between these frequencies and our highest observed
frequency at 200~Hz. These results support the theoretical conclusion by \citet{2014MNRAS.440.3613H}
that the scheme proposed by \citet{2013ApJ...763L..44L} is insensitive to dispersion measure.

\subsection{Origins of Increased Spectral Power Below 10~Hz}\label{sec5:subHz}

We associate the increased spectral power indicated by the dotted and dashed curves in Fig.~\ref{fig2}
with scintillation caused by the interplanetary medium. We associate the dot-dashed curves with ionospheric
scintillation. We justify these associations as follows.

The dotted and dashed curves in Fig.~\ref{fig2} are consistent with power spectra for interplanetary scintillation
reported throughout the literature, for example \citet{2015SoPh..290.2539M} who observed 3C~48 at 140~MHz and 327~MHz
during the current peak of Solar Cycle 24 at similar elongations to our data. The frequency at the knee where
the curves change slope corresponds to the Fresnel filter frequency, $f_F$, which is in turn proportional to
the velocity of the solar wind \citep[e.g.][]{1990MNRAS.244..691M}. The solar-wind velocity projected onto the plane
perpendicular to the line of sight can be estimated as $V_\perp \approx f_F \sqrt{\pi \, \lambda \, A \cos(\epsilon)}$
where $\lambda$ is the observing wavelength, $A$ is 1~AU, and $\epsilon$ is solar elongation. Taking $f_F=1.5$~Hz
from \S~\ref{sec4} and $\epsilon=23^\circ$ from Table~\ref{tbl-obs} we estimate $V_\perp\approx410$~km~s$^{-1}$ at the
2014 May 15 epoch. This is consistent with the $\sim400$~km~s$^{-1}$ velocity of the slow solar wind
\citep[e.g.][]{2005JGRA..110.7109F}. The data at the 2014 Apr 11 epoch (dotted curve) exhibit excess power compared
to 2014 May 15 (dashed curve), appearing shifted to the right with $f_F\approx2.3$~Hz and thus $V_\perp\approx570$~km~s$^{-1}$.
The cause of this shift is likely the large coronal mass ejection CME52 that departed the Sun at 2014
Apr 8 23:12 UTC in the direction of 3C~84 with median velocity 488~km~s$^{-1}$ and velocity range 230--900~km~s$^{-1}$,
as reported in the CACTus\footnote{http://sidc.oma.be/cactus/} quicklook catalog for the LASCO instrument onboard
the {\it Solar and Heliospheric Observatory} \citep{2004A&A...425.1097R}. For reference, ejecta traveling at
600~km~s$^{-1}$ oriented perpedicular to the line of sight to the sun would have traversed 3C~84's position
at the time of our GBT observations.

Interplanetary scintillation is unlikely to be the origin of the excess power observed at the GBT reference
position on 2014 Apr 11, indicated by the dot-dashed curve in Fig.~\ref{fig2}. This is because we expect Galactic
emission at the reference position to be mostly smooth on the angular scales {\footnotesize $\lesssim$}$1^{\prime\prime}$
required for interplanetary scintillation \citep{1964Natur.203.1214H}. Instead, we attribute the excess power
to ionospheric scintillation where the critical angular scale is {\footnotesize $\lesssim$}$2^{\prime}$
(given by the Fresnel scale $\sqrt{\lambda d}$ at the height of the ionosphere $d\approx400$~km). The
dot-dashed curve is consistent with power spectra for ionospheric scintillation reported in the literature,
for example \citet{fang} who present spectra with $f_F\approx0.14$~Hz and power law slope $\approx -1.8$
from observations of {\it Global Positioning System} satellites. Their data were obtained in 2008--2009
during solar minimum. Our higher Fresnel frquency at 0.25~Hz is likely due to increased solar activity in
2014--2015. We also attribute the excess power observed on 2015 Jan 2 to ionospheric scintillation.
Interplanetary scintillation is insufficient to explain this power due to the large solar elongation of
3C~84 and the expected $\epsilon^{-4}$ dependence for interplanetary scattering power \citep{1978ApJ...220..346A}.
Finally, we tentatively note that the Fresnel frequency associated with the GBT reference position data
may be slightly higher than for the 2015 VLA data. This could possibly be due to the class C9.4 solar
flare emitted 3 hours prior to our GBT observation, as cataloged by the X-ray Sensor on board the
{\it GOES--15} satellite\footnote{http://www.ngdc.noaa.gov/stp/satellite/goes/} \citep{1994SoPh..154..275G}.
For comparison, no flares greater than class C2.1 were observed in the 24~h preceding our 2015 observation.

\section{Conclusions}\label{sec6}

We examined high cadence VLA and GBT observations of 3C~84 at 1.7~GHz for signatures of intergalactic
dispersion predicted by \citet{2013ApJ...763L..44L}. Our data constrain the ratio between standard deviation
and mean intensity for 3C~84 to a conservative upper limit of 0.05\% over timescales spanning 0.1--200~Hz.
This limit is 3 orders of magnitude less than predicted, specifically within the 1--4~Hz and 10--40~Hz
ranges appropriate for 3C~84's DM and our two observational setups.

Our finding is consistent with the theoretical analysis of \citet{2014MNRAS.440.3613H}, who concluded
that the scheme proposed by \citet{2013ApJ...763L..44L} \citep[and similarly by][]{2013MNRAS.433.2275L}
is insensitive to dispersion measure. We note that, had the theory been correct, it is plausible that
indicative signatures could have escaped detection throughout historical radio observations due to the
subtle predicted dependencies on bandwidth and source DM \citep[e.g. the dedicated high cadence search
for anomalous variability from 3C~84 by][]{2015arXiv150600055H}.

We investigated the origins of increased temporal spectral power below 10~Hz in our data. We found
contributions consistent with the slow solar wind, a coronal mass ejection, the ionosphere, and
possibly the excited ionosphere perturbed by a solar flare.

Our variability limit for 3C~84 demonstrates that signal paths through the VLA's WIDAR correlator and
the GBT's VEGAS backend are of high fidelity. We limit the presence of any contaminating artifacts to
the equivalent of a sinusoid with amplitude $<10$~mJy and frequency in the range 0.1--200~Hz for
observations at $>2.5$~ms cadence.

Finally, to our knowledge, we have presented the highest time resolution search for intrinsic radio
variability from a supermassive black hole to date. While intrinsic variability is not expected from
3C~84 due to its $\sim1$~hour light crossing timescale, we are unaware of any existing data that
could rule out such behavior. Historically, astronomers didn't expect interstellar scintillation,
pulsars, or gamma-ray bursts, either.

\acknowledgments

The National Radio Astronomy Observatory is a facility of the National Science
Foundation operated under cooperative agreement by Associated Universities, Inc.
CAH acknowledges support from a Jansky Postdoctoral Fellowship from NRAO.
We thank the following for support and helpful feedback: Tim Bastian,
Walter Brisken, Paul Demorest, Vivek Dhawan, Chris Hirata, Amanda Kepley,
Brian Mason, Toney Minter, Rick Perley, Richard Prestage, Alan Roy, and Craig Walker.
We thank the anonymous referee for comments that led to the improvement of this work.

{\it Facilities:} \facility{VLA}, \facility{GBT}, \facility{SOHO}, \facility{GOES}.

\end{document}